\documentclass[12pt,a4paper]{article} 
\usepackage{epsfig}
\usepackage{graphics}
\usepackage{subfigure}
\usepackage{a4wide}
\usepackage{amsmath}

\title{Topological Q-Solitons}
 \author{R.\ S.\ Ward\footnote{email: richard.ward@durham.ac.uk}
 \bigskip
\\Department of Mathematical Sciences,  \\ University of
Durham, \\Durham DH1 3LE}

\newcommand{\half}{{\scriptstyle\frac{1}{2}}}
\newcommand{\quar}{{\scriptstyle\frac{1}{4}}}
\newcommand{\RR}{{\bf R}}

\newcommand{\vphi}{\vec\phi}
\newcommand{\pa}{\partial}
\newcommand{\ee}{{\rm e}}
\newcommand{\ii}{{\rm i}}
\renewcommand{\o}{\omega}

\begin{document}
\maketitle \abstract{\noindent
Static topologically-nontrivial configurations in sigma-models, for spatial
dimension $D\geq2$, are unstable. The question addressed here is whether
such sigma-model solitons can be stabilized by steady rotation in internal
space; that is, rotation in a global SO(2) symmetry.  This is the mechanism
which stabilizes Q-balls (non-topological solitons).  The conclusion is that
the Q-mechanism can stabilize topological solitons in $D=2$ spatial
dimensions, but not for $D=3$.
}
\newpage

\section{Introduction}

Topological solitons (stable, localized, topologically non-trivial
solutions in a field theory) have long been of great interest, both
for their mathematical properties and for their
applications in many areas of physics.
In any system admitting such solitons, the non-trivial topology is
not sufficient to ensure that the solitons are stable.  One obvious
reason for this is that topology is, by definition, a non-metric
structure, and so it cannot determine the size of the solitons; 
for that, one needs to balance the forces acting on the soliton in such
a way that it has a preferred size.  Recall, for example, the
O($n$) sigma model in $D+1$ dimensions (with trivial boundary condition
at spatial infinity), which is the subject of this article.
This system admits topological configurations (textures)
whenever the homotopy group $\pi_D(S^{n-1})$ is non-trivial; in
particular, for $(D,n)$ equal to $(2,3)$, $(3,3)$ or $(3,4)$.
If $D=3$, then solitons tend to shrink --- in the pure sigma model,
there are no static solutions.  In the $D=2$ case, there are static
solutions, for example the Belavin-Polyakov solitons \cite{BP75} in the
O(3) system; but these are unstable \cite{LPZ90,PZ96}.

A soliton can always be prevented (or rather discouraged) from
spreading out by the addition, if necessary,
of a potential (a term involving only the field, and not its gradient).
In order to stabilize the soliton size, we also need to introduce
something which prevents it from shrinking.  There are several
possibilities for such an anti-shrinking mechanism: for example, a Skyrme
term involving four (or more) powers of the field gradient; or
a gauge field suitably coupled to the sigma-field; or
periodic time-dependence (rotation in an internal space).
This third possibility also underlies non-topological solitons
(Q-balls).  In this paper, we investigate to what extent the Q-ball
mechanism is effective at stabilizing
topological sigma-model solitons.  We shall see
that stationary topological Q-solitons exist in $D=2$ spatial dimensions,
but not for $D=3$.  This result is analogous to that for the
Landau-Lifshitz equation (Heisenberg model of ferromagnetism),
as one might have surmised since the static Landau-Lifshitz system is
identical to the static sigma model.

\section{Q-Solitons in the O(3) Sigma Model}

The O(3) sigma model involves a scalar field taking values on $S^2$;
this field can
be represented as a unit 3-vector $\vphi=(\phi_1,\phi_2,\phi_3)$
with $\vphi\cdot\vphi=1$.  The Lagrangian is
\begin{equation}\label{Lag}
 {\cal L} = \half(\pa_\mu\vphi)\cdot(\pa^\mu\vphi)
               - V(\phi_3),
\end{equation}
where $V$ is some potential function (which, for simplicity, we take
to depend only on $\phi_3$).  The space-time coordinates are
$x^\mu = (t,x^j)$, with $j=1,\ldots,D$.  The system has a global
SO(3) symmetry which is broken, by the potential term, to SO(2).
This SO(2)
acts only on $\phi_1$ and $\phi_2$, namely by changing the phase of
$\phi := \phi_1+\ii\phi_2$. The corresponding conserved quantity is
\begin{equation}\label{Q1}
   Q = \int{\rm Im}(\dot\phi\bar\phi)\,d^Dx.
\end{equation}
Minimizing the energy of a configuration subject to $Q$ being fixed
implies \cite{LP92}, in particular, that $\vphi$ has the form
\begin{equation}\label{QBF}
   \phi(t,x^j)=\psi(x^j)\ee^{\ii\o t},\quad \phi_3=\psi_3(x^j),
\end{equation}
with $|\psi|^2+(\psi_3)^2=1$.  Without loss of generality,
we shall assume that $\o\geq0$.  Note that $Q=\o I$,
where $I = \int|\psi|^2\,d^Dx$.  The energy of a configuration of
the form (\ref{QBF}) is $E = E_d + E_k + E_p$, where
\begin{eqnarray*}
   E_d &=& \half\int\left[|\pa_j\psi|^2+
               (\pa_j\psi_3)^2\right]\,d^Dx, \cr
   E_k &=& \half I\o^2 = \half Q^2/I, \cr
   E_p &=& \int V(\psi_3)\,d^Dx.
\end{eqnarray*}
The boundary condition is $\psi_3\to1$ as $r\to\infty$; so we need
$V(1)=0$.

A stationary Q-lump is a critical point of the energy functional
$E[\vec\psi]$, subject to Q having some fixed value.  Such a
Q-lump is (classically) stable if this critical point is
a local minimum of $E$.  The usual (Derrick) scaling argument
shows that any stationary Q-lump must satisfy
\begin{equation}\label{scaling}
   (2-D)E_d - D E_p + D E_k = 0.
\end{equation}
Let the positive constant $m$ be defined by $V'(1)=-m^2$; in other
words, $V(\psi_3) \approx m^2(1-\psi_3) \approx \half m^2|\psi|^2$ for
$\psi_3\approx1$.  Then, near spatial infinity, the Euler-Lagrange
equations corresponding to $E$ imply that
\[
  \nabla^2\psi-(m^2-\o^2)\psi=0.
\]
So in order to satisfy the
boundary condition $\psi\to0$ as $r\to\infty$, we need $\o\leq m$.
The solitons are exponentially localized if $\o < m$, but
less-localized solitons with $\o=m$ may also exist.

The parameter $m$ is part of the specification of the system, and
the parameter $Q$ is set by the initial data.  Each of these two
parameters has dimensions; the combination $Qm^{D-1}$ is dimensionless,
whereas the combination $(Q/m)^{1/D}$ has dimension of length, and
determines the size of the soliton.
Configurations in this system are classified topologically by their
topological charge $N$ (an integer); if $D=2$, then $N$ is the winding
number, while if $D=3$, then $N$ is the Hopf number.  Let $E(N,Q)$
denote the energy of a configuration (or rather, of data $\vphi,\vphi_t$)
with topological charge $N$ and Noether charge $Q$.

Let $V_0$ be the normalized potential function $V_0 := 2V/m^2$.  Note
that $V_0(\psi_3)/|\psi|^2 \to 1$ as $\psi_3\to1$.
It is clear from (\ref{scaling}) that if $V_0\geq|\psi|^2$ everywhere,
with $V_0>|\psi|^2$ somewhere, then there can be no solution.
So the constant $K$ defined by
$K = {\rm min}\left[V_0(\psi_3)/|\psi|^2\right]$ should satisfy
$K\leq1$.  It then follows that
\begin{equation}\label{omega-inequality}
 E_p = m^2\int V_0 \geq m^2KI = (Km^2/\o^2)E_k \geq (Km^2/\o^2)E_p,
\end{equation}
where the final inequality comes from (\ref{scaling}).
As a consequence, we have
\begin{equation}\label{omega}
   K m^2 \leq \o^2 \leq m^2.
\end{equation}
In the $D=3$ case, the first inequality is strict:
$K m^2 < \o^2 \leq m^2$.

In two spatial dimensions, it is possible to have $K=1$, which corresponds
to the choice $V(\psi_3)=\half m^2(1-\psi_3^2)$. So here $\o=m$. This system
\cite{L91},
and generalizations in which the target space is some other K\"ahler
manifold, arise naturally by dimensional reduction from `pure' sigma
models in one dimension higher \cite{A92}.  The energy satisfies a
Bogomolny bound $E(N,Q) \geq 4\pi N + mQ$, which (for $N\geq2$) can
be saturated: for each value of $N\geq2$ and $Q$, there is an explicit
family ${\cal M}$ of stationary multi-soliton solutions such that
$E(N,Q)= 4\pi N + mQ$.  There is no force between the individual
solitons: in particular, the total energy
has the additive property $E(N_1,Q_1)+E(N_2,Q_2) = E(N_1+N_2,Q_1+Q_2)$.
One may use moduli-space methods (as was done in \cite{W85,L90,S91}
for other sigma-model systems) to investigate the scattering of moving
solitons \cite{L91}; this involves finite-dimensional mechanics on
${\cal M}$.  The dynamics turns out to be rather exotic (as is
also the case \cite{BS00,AKPF00,MV00} for non-topological Q-balls).
The solitons are only polynomially localized, and the non-existence of
an $N=1$ soliton is related to this; an $N=1$ configuration tends to
shrink in size, and there is no stationary $N=1$ solution.

On the other hand, if $K<1$, then 1-solitons can exist. Different choices
of $V$ (having $K<1$) seem to lead to similar behaviour, but this
has yet to be fully investigated; in what follows, we take
$V_0=\half(1-\psi_3^4)$, so $K=\half$.  Let us consider, first,
the thin-wall limit \cite{C85}, where
$Qm^{D-1}\gg1$.  In this limit, the (bulk) contributions $E_p$ and $E_k$
to the energy are very much greater than the (surface) contribution
$E_d$.  So the energy is approximately
$E\approx\quar m^2\int(1-\psi_3^4) + \half Q^2/\int(1-\psi_3^2)$.
Without loss of generality, we may assume that $\psi_3=1$ outside of some
compact set.  So space is partitioned into three regions: one (with
infinite volume) where $\psi_3=1$, the second (with volume $A$) where
$|\psi_3|\neq1$, and the third (with volume $B$) where $\psi_3=-1$.
Note that $E$ depends on $A$ (and on the value that $\psi_3$ takes
on $A$), but not on $B$, and that
\[
  \frac{\delta E}{\delta\psi_3} = (\o^2-m^2\psi_3^2)\psi_3.
\]
So for fixed $A$, the function $E$ has a minimum for $\psi_3=0$
($\psi_3=\pm\o/m$ are local maxima).  Hence we should set $\psi_3=0$
on $A$, and the energy becomes
\[
  E = \half Q^2/A + \quar m^2 A.
\]
Thus for a given value of $Q$, the energy has the minimum value
$E_{{\rm min}}=mQ/\sqrt{2}$ when $A=Q\sqrt{2}/m$.  Note that
$\o=m/\sqrt{2}$, at the lower end of its allowed range (\ref{omega}).

\begin{figure}[htb]
\begin{center}
{
\includegraphics[scale=0.5]{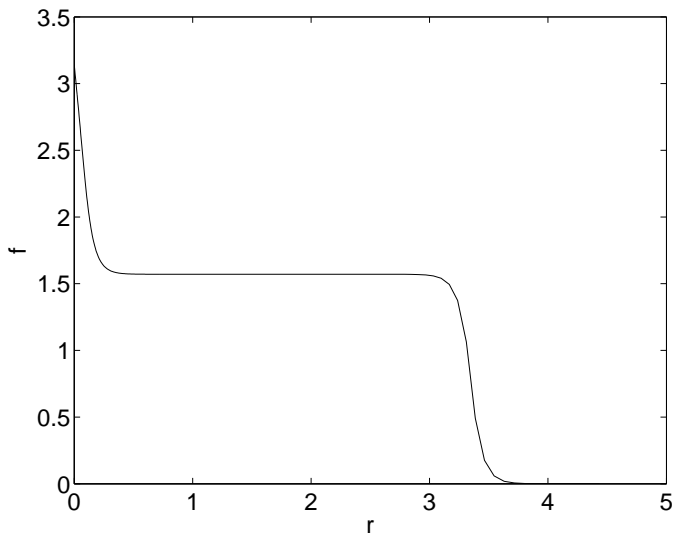}
\caption{The profile function $f(r)$ for the 1-soliton solution on the
        plane, with $m=20$ and $Q=160\pi$.\label{fig1}}
}
\end{center}
\end{figure}

To make further progress, we need to include the effect of surface
tension, in other words include the term $E_d$.  Let us consider, first,
the planar case $D=2$.  For simplicity, we assume rotational symmetry
about a point in the plane: the field is taken to have the form
$\psi=\sin(f)\exp(\ii N\theta)$ and $\psi_3=\cos(f)$, where $f=f(r)$.
The boundary conditions are $f(0)=\pi$ and $f(\infty)=0$, and $N$ is the
topological charge.  The energy functional $E = E_d + E_k + E_p$
was minimized numerically, for various values of $m$, $Q$ and $N$. 
The term $E_k$ was used in the form
$E_k = Q^2/I$, so that one can minimize while keeping $Q$ fixed;
the quantity $\omega$ does not enter explicitly, but can be derived
(via the formula $\o=Q/I$) once the minimum has been found.  In each case
that was investigated, a smooth minimum was reached.  For $N=1$ and
$Qm=3200\pi$ (close to the thin-wall limit), the profile function
$f(r)$ is plotted in Fig~\ref{fig1}.  We see that there is a region
around $r=0$ where $f(r)$ drops rapidly from $\pi$ to $\pi/2$
(the term $E_d$ prevents this region from shrinking further in size);
and then a region (corresponding to $A$ in the argument above) where
$f=\pi/2\Leftrightarrow\psi_3=0$.  Outside of this region, the field
takes on its asymptotic value $f=0\Leftrightarrow\psi_3=1$. 

\begin{figure}[htb]
\begin{center}
{
\subfigure{
\includegraphics[scale=0.4]{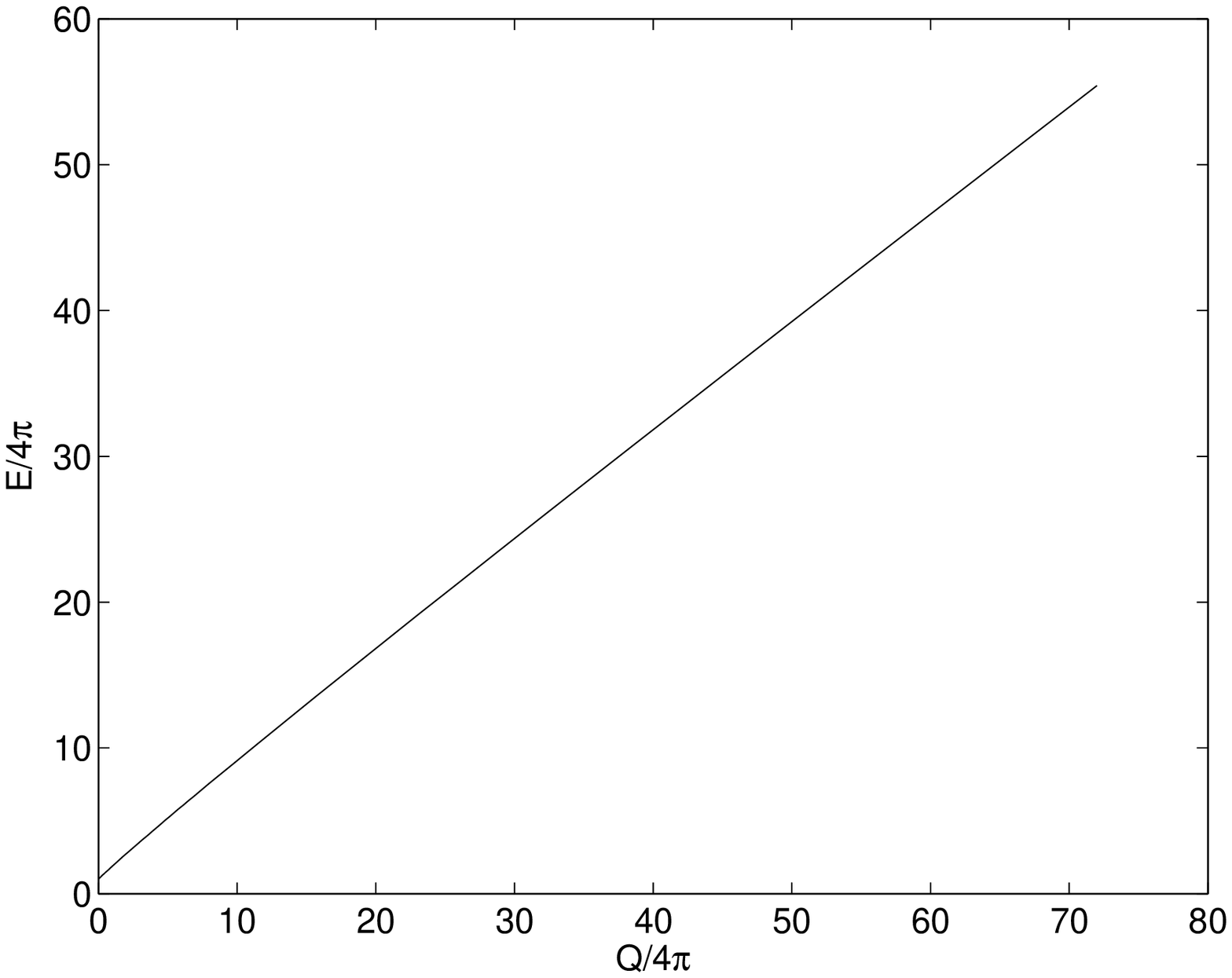}
}
\quad
\subfigure{
\includegraphics[scale=0.4]{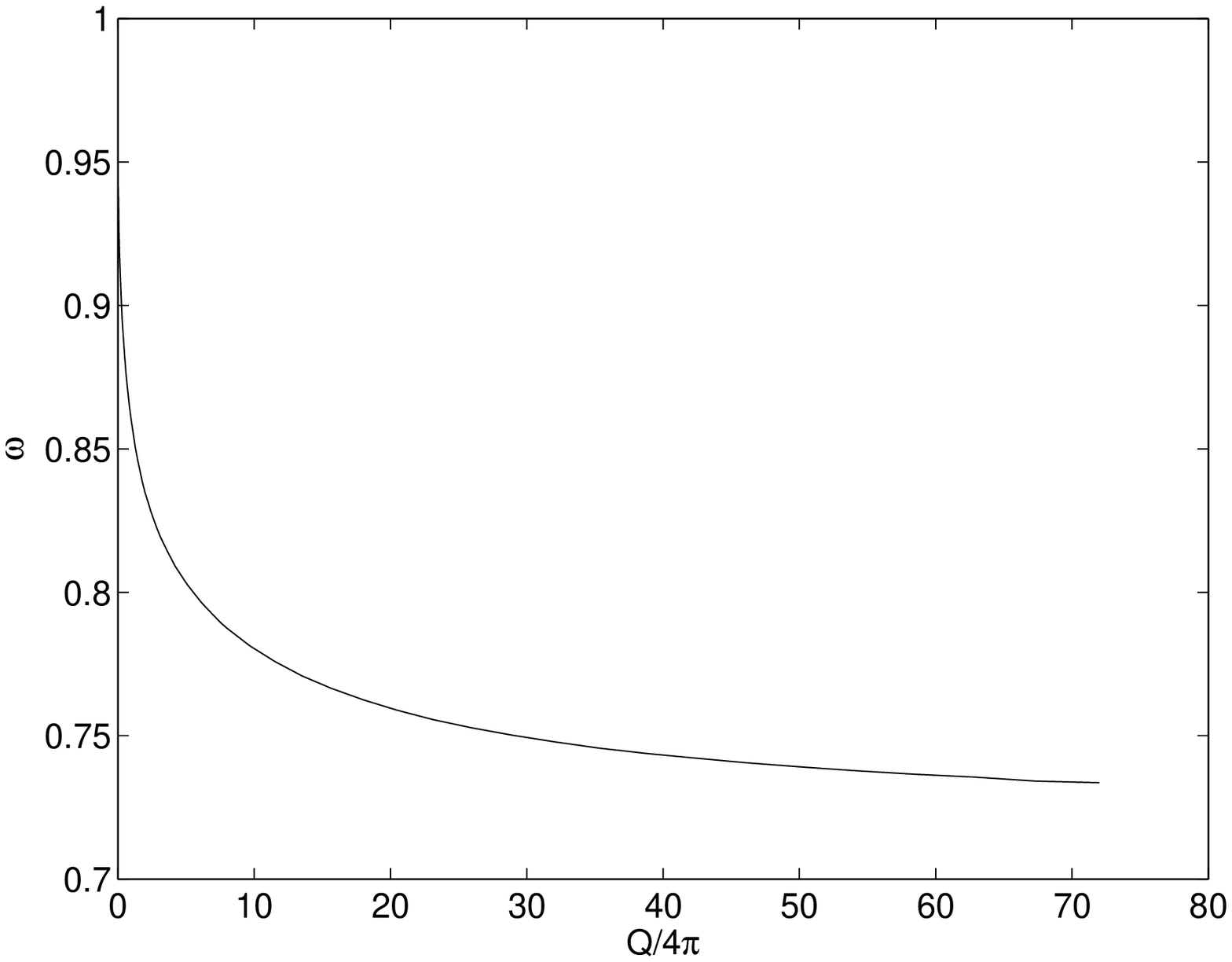}
}
\caption{The energy $E$ and angular frequency $\o$ of the 1-soliton,
  as functions of $Q$.  \label{fig2}}
}
\end{center}
\end{figure}

Fig~\ref{fig2} displays results for $N=1$, $m=1$ and a range of values
of $Q$.  We see that
$E$ is very close to being linear in $Q$ (recall that in the Bogomolny
case, it is exactly linear): to a very good approximation, we have
$E = 4\pi + 3Q/4$.  As for $\o$, we know from (\ref{omega}) that $\o$
has to be in the range $1/\sqrt{2} < \o < 1$, and we see from the figure
that this is so; furthermore, $\o\to1/\sqrt{2}$ as $Q\to\infty$ (the
thin-wall limit \cite{C85}), while $\o\to1$ as $Q\to0$ (the
thick-wall limit \cite{K97}).

In the Bogomolny case \cite{L91} mentioned previously, the energy
$E(N,Q)$ of a stationary soliton has the feature that, for a given $Q$,
the quantity $E_N:=(4\pi N)^{-1}E(N,NQ)$ is independent of $N$:
in fact, $E_N=1+mQ/4\pi$.  This corresponds to the fact that in this
Bogomolny-type system, there is no force between stationary solitons.
For the potential $V=\quar m^2(1-\psi_3^4)$, however, there are such
forces. This can be seen by examining $E_N$ for fixed $Q=4\pi$ and
for various values of $N$.  Table~\ref{tab1}
shows the results for $1\leq N \leq7$.  The energy density of the
1-soliton is peaked at the point $r=0$, whereas that for the
rotationally-symmetric $N$-soliton is peaked on a ring.  Note that
$E_N$ is a decreasing function of $N$, which suggests that the
this $N$-soliton is stable against breakup into
solitons of lower topological charge.  But this remains to be checked;
in particular, one should investigate the vibrational modes about these
rotationally-symmetric solutions.
\begin{table}[thb]
 \begin{center}
  \begin{tabular}{lccccccc}\hline
    $N$ & 1 & 2 & 3 & 4 & 5 & 6 & 7 \\
    \hline
      $E_N$  & 1.888 & 1.840 & 1.826 & 1.820 & 1.817 & 1.816 & 1.815 \\
      $\o/m$ & 0.859 & 0.826 & 0.817 & 0.814 & 0.882 & 0.811 & 0.810 \\
  \end{tabular}
  \caption{Energy $E_N=E(N,NQ)/(4\pi N)$ and frequency $\o/m$,
     for $m=1$ and $Q=4\pi$}\label{tab1}
 \end{center}
\end{table}

Finally, let us turn to the case of $D=3$ spatial dimensions.
Configurations with nonzero Hopf number $N$ look like closed loops,
which may be linked or knotted.  Static knot-solitons are have been studied
in the Faddeev-Skyrme system, where a Skyrme term is added to the
Lagrangian: this extra term stabilizes the solitons, which would otherwise
shrink \cite{FN97,GH97,BS99,W00}.   The question here is whether there
exist stationary Hopf solitons which are stabilized by internal rotation
rather than by a Skyrme term.
The answer to this question appears to be negative; what happens
is as follows.  Consider an $N=1$ configuration.  Typically, $\psi_3=1$
at spatial infinity and on a curve which extends to infinity; let
us visualize this curve as the $z$-axis.  Secondly, $\psi_3=-1$ on a closed
loop $L$ around the $z$-axis.  Finally, $\psi_3=0$ on a torus around the
$z$-axis, with the loop $L$ in its interior.  So the region $A$ in the
previous thin-wall analysis is a thickened torus (with the $B$-region
in its interior) resembling a closed string.   Numerical experiments
indicate that,
roughly speaking, the Q-effect supports the thickness of the string,
but not its length; the string has a tension which causes its length
to shrink.  So the configuration
collapses, and there is no stationary Hopf soliton in this system.
In principle it remains a possibility that for some potential function
$V$, some value of $Q$, and some nonzero value of the Hopf number $N$,
there might exist a stationary solution; but this seems rather unlikely.

Another way of viewing the situation is as follows. The Q-mechanism provides
a lower bound on the quantity $I$ (since $\omega$ is bounded above,
and $Q=I\omega$ is fixed); this in turn means that the volume
of the soliton is bounded below.  But surface tension then acts to make
the soliton spherical.  So we are led to the following conjecture:
any stationary Q-ball (whether topological or not) with $\omega<m$,
in $D$ spatial dimensions, has O($D$) symmetry.  (In the Bogomolny case,
where $\omega=m$, rotational symmetry is not essential \cite{L91}.)
The instability of Hopf Q-solitons is an immediate consequence of this
conjecture, since O(3) symmetry implies that the Hopf number is zero.

\section{The O(4) Sigma Model in 3+1 Dimensions}

In this section, we investigate the analogous problem for the
O(4) sigma model in three space dimensions.
The details of the system are similar to those of the 
previous section.  The field takes values on $S^3$, and is
represented as a unit 4-vector $\vphi=(\phi_0,\phi_1,\phi_2,\phi_3)$.
The Lagrangian is (\ref{Lag}), as before, but with the potential
being allowed to depend on $\phi_0$ and $\phi_3$: $V=V(\phi_0,\phi_3)$.
In general, this breaks the global O(4) symmetry to O(2), the subgroup
which rotates $\phi_1$ and $\phi_2$, leaving $\phi_0$ and $\phi_3$ fixed.
The expressions for $E_d$, $E_k$ and $E_p$ are the same as before,
with $D=3$,
except that in $E_d$ there is an extra term involving $(\pa_j\psi_0)^2$.

We take the boundary condition to be $\psi_0\to1$ as $r\to\infty$
in $\RR^3$.  The mass $m$ is defined by
$V(\psi_0,\psi_3)\approx m^2(1-\psi_0)\approx\half m^2(|\psi|^2+\psi_3^2)$
for $\psi_0\approx1$; the function $V_0$ and the constant $K$ are defined
as before.  The virial relation from (\ref{scaling}) with $D=3$
holds as before, as does the inequality $K m^2 < \o^2 \leq m^2$. 

In \cite{OS90}, this system was studied, with the potential
$V_0=2(1-\psi_0)$.  Since in that case we have $K=1$, no soliton solution
can exist. The authors of \cite{OS90} reach this conclusion for the
topologically trivial case $N=0$; they report numerical evidence for a
non-trivial solution with $N=1$, but this cannot be correct.

In order to allow the possibility of non-trivial solutions, we need a
potential $V_0$ which has $K<1$; for what follows, we shall take
$V_0 = \half(1-\psi_0^4)$ as in the previous section.  Then there are
solutions, but it appears that they all have trivial topology ($N=0$).
One way to see what happens is to consider the thin-wall limit, where
$m^2Q\gg1$.  So the energy is approximately
$E\approx\quar m^2\int(1-\psi_0^4) + \half Q^2/\int|\psi|^2$;
and the corresponding variational equations are
\begin{equation} \label{thinwall}
  \xi\psi_0-m^2\psi_0^3=0, \quad
  \xi\psi-\o^2\psi=0, \quad
  \xi\psi_3=0,
\end{equation}
where $\xi := m^2\psi_0^4+\o^2|\psi|^2$ (the $\xi$-term arises from
enforcing the constraint $\vphi\cdot\vphi=1$).  These equations
(\ref{thinwall}) have a number of solutions, namely:
\begin{itemize}
  \item $\psi_0=0$, $\psi=0$, $\psi_3=\pm1$;
  \item $\psi=0$, $\psi_3=0$, $\psi_0=\pm1$;
  \item $\psi_0=\psi_3=0$, $|\psi|^2=1$;
  \item $\psi_3=0$, $\psi_0=\pm\o/m$, $|\psi|^2=1-\o^2/m^2$.
\end{itemize}
So to construct a mimimum-energy configuration, we must partition space
$\RR^3$ into regions (separated by infinitesimally-thin walls), on each
of which one of these relations holds.  It is clear that regions on
which $\psi_3\neq0$ contribute only to $E_p$, and that we can reduce
the total energy by instead setting $\psi_3=0$, $\psi_0=1$ on these
regions.  In other words, $\psi_3$ `collapses' to zero, and is replaced
by $\psi_0$.

This is exactly what one sees in numerical simulations.  For example,
we may start with the O(3)-symmetric `hedgehog' ansatz 
\begin{equation}\label{hedgehog}
   \psi_0 + \ii\psi_j\sigma_j = \exp\left[\ii f(r)\,x^j\sigma_j/r\right];
\end{equation}
here $\sigma_j$ denotes the Pauli matrices, and the profile function $f(r)$
satisfies the usual boundary conditions $f(0)=\pi$, $f(\infty)=0$. The
winding number is $N=1$. Note that $\psi=\psi_1+\ii\psi_2$ vanishes on the
$x^3$-axis, and that $\psi_0(0)=-1$.  If we now relax the configuration
by flowing down the energy gradient, then $\psi_3$ approaches zero
everywhere except at the single point $r=0$; in other words, there
is no continuous minimum in this
topological class.  By contrast, there is a smooth minimum which has
$\psi_3\equiv0$: this is topologically trivial, and is essentially
a standard (nontopological) Q-ball.

\section{Concluding Remarks}

We begin with a few remarks on the similarities with stationary topological
soliton solutions of the Landau-Lifshitz equation
\begin{equation} \label{LLeqn}
 \frac{\pa\vphi}{\pa t} = - \vphi\times\frac{\delta E}{\delta\vphi}.
\end{equation}
Here $\vphi$ is a unit 3-vector representing the local orientation of
magnetization, and the energy $E$ is given by
\begin{equation}\label{LLenergy}
E=\int \left[\half(\pa_j\vphi)\cdot(\pa_j\vphi)+U(\phi_3)\right]\,d^Dx.
\end{equation}
A typical choice for the function $U$ is $U=A(1-\phi_3^2)$, where $A$
is a constant; this corresponds to an easy-axis anisotropy.  The boundary
condition is $\phi_3\to1$ as $r\to\infty$.  The total magnetization
\begin{equation}\label{LLmag}
  M = \int (1-\phi_3)\, d^Dx
\end{equation}
is a conserved quantity.  It is clear from scaling that the only static
solutions are the Belavin-Polyakov solitons \cite{BP75} in spatial
dimension $D=2$, with $U\equiv0$.  But by allowing time dependence,
more solutions are possible.  In particular, we may allow periodic
time dependence, and look for stationary solutions such that
\begin{equation}\label{LLstat}
   \phi_1+\ii\phi_2=\psi(x^j)\ee^{\ii\nu t},\quad \phi_3=\psi_3(x^j).
\end{equation}
With this ansatz, the Landau-Lifshitz equation (\ref{LLeqn}) is
equivalent to
\begin{equation} \label{LLstateqn}
  \frac{\delta}{\delta\vphi} \left(E-\half\nu M\right) = 0.
\end{equation}
We may think of the solutions as critical points of $E$, subject to the
constraint that $M$ has a given value \cite{TW77}.
The simplest example occurs if $U=\half\nu(1-\psi_3)$, for then the
functional appearing in (\ref{LLstateqn}) consists of only the gradient
term, and the Belavin-Polyakov solitons are solutions (in $D=2$); these
correspond, of course, to the Bogomolny-type Q-lump solitons \cite{L91}.
The analysis of stationary topological Landau-Lifshitz solitons leads to
rather similar results as for Q-solitons (although the dynamics of moving
solitons is quite different).  In $D=2$, there are topological solutions
(called magnetic bubbles --- see \cite{KIK90} for a review); a
single soliton is pinned in space, and cannot move \cite{PT91}.
In $D=3$, on the other hand, there are no stationary Hopf solitons
\cite{P93}; however, such solitons can be stabilized by allowing them
to move at constant velocity \cite{P93,C99}.

Returning to sigma-model dynamics, we have seen that the Q-mechanism
stabilizes topological solitons in $D=2$ spatial dimensions, but not in
$D=3$.  Stabilizing vortex rings (Hopf textures) in $D=3$ is particularly
difficult, since there are two length-scales (the length of the loop
and its width), each of which has to be fixed.  The Q-effect can
stabilize the latter, but not the former.  One does get stable loops
in systems with a Skyrme term \cite{FN97,GH97}, and also in systems with
a magnetic field sufficiently strongly coupled to the scalar field
(minimal coupling is not enough) \cite{W02}.  But in the basic versions
of each of these systems, there is only one length-scale; and
so the length of the loop is of the same order as (and only slightly
greater than) its thickness.  It remains an open question as to
whether there is a system admitting a stable Hopf soliton in which
the two length-scales are significantly different.

A sigma-model soliton in $D=2$ can be thought of as a (straight) sigma-model
string in three spatial dimensions.  So, for example, the Q-stabilized
solitons discussed in this paper may find application as cosmic strings.
Given an appropriate potential $V$, long strings with internal rotational
energy will be stable, although closed loops will eventually shrink and decay.
In this connection, it is worth recalling that, on a cosmological scale,
both the width and the length of sigma-model strings are stabilized by
cosmological expansion; but stabilizing the length requires a greater rate
of expansion than stabilizing the width \cite{W02b}.


\begin{thebibliography}{99}

\bibitem{BP75}
A~Belavin and A~Polyakov, Metastable states of two-dimensional
 isotropic ferromagnets.  {\it JETP~Lett} {\bf22} (1975) 245--247.

\bibitem{LPZ90}
R~A~Leese, M~Peyrard and W~J~Zakrzewski, Soliton stability in the
  O(3) sigma model in (2+1) dimensions.
  {\it Nonlinearity} {\bf3} (1990) 387--412.

\bibitem{PZ96}
B Piette and W~J~Zakrzewski, Shrinking of solitons in the
  (2+1)-dimensional $S^2$ sigma model.
  {\it Nonlinearity} {\bf9} (1996) 897--910.

\bibitem{LP92}
T~D~Lee and Y~Pang, Nontopological solitons.
  {\it Physics~Reports} {\bf221} (1992) 251--350.

\bibitem{L91} 
R~Leese, Q-lumps and their interactions.
  {\it Nucl Phys B} {\bf366} (1991) 283--311.

\bibitem{A92} 
E~Abraham, Non-linear sigma models and their Q-lump solutions.
  {\it Phys Lett B} {\bf278} (1992) 291--296.

\bibitem{W85}
R S Ward, Slowly-moving lumps in the CP$^1$ model in (2+1)
  dimensions.  {\it Phys~Lett~B} {\bf158} (1985) 424--428.

\bibitem{L90}
R A Leese, Low-energy scattering of solitons in the CP$^1$ model.
         {\it Nucl~Phys~B} {\bf344} (1990) 33--72.

\bibitem{S91}
P~M~Sutcliffe, The interaction of Skyrme-like lumps in (2+1)
  dimensions.  {\it Nonlinearity} {\bf4} (1991) 1109--1121.

\bibitem{BS00} 
R~A~Battye and P~M~Sutcliffe, Q-ball dynamics.
  {\it Nucl Phys B} {\bf590} (2000) 329--363.

\bibitem{AKPF00}
M~Axenides, S~Komineas, L~Perivolaropoulos and M~Floratos,
  Dynamics of nontopological solitons --- Q balls.
  {\it Phys~Rev~D} {\bf61} (2000) 085006.

\bibitem{MV00} 
T~Multam\"aki and I~Vilja, Analytical and numerical properties
  of Q-balls. {\it Nucl~Phys~B} {\bf574} (2000) 130--152.

\bibitem{C85} 
S~Coleman, Q-balls. {\it Nucl Phys B} {\bf262} (1985) 263--283.

\bibitem{K97} 
A~Kusenko, Small Q-balls.
  {\it Phys Lett B} {\bf404} (1997) 285--

\bibitem{FN97}
L~Faddeev and A~J~Niemi, Stable knot-like structures in classical
  field theory.  {\it Nature} {\bf387} (1997) 58--61.

\bibitem{GH97} 
J~Gladikowski and M~Hellmund, Static solitons with nonzero Hopf
  number. {\it Phys~Rev~D} {\bf56} (1997) 5194--5199.

\bibitem{BS99}
R~A~Battye and P~M~Sutcliffe, Solitons, Links and Knots.
    {\it Proc Roy Soc Lond A} {\bf455} (1999) 4305--4331.

\bibitem{W00}
R~S~Ward, The interaction of two Hopf solitons.
       {\it Phys Lett B} {\bf473} (2000) 291--296.

\bibitem{OS90} 
H~Otsu and T~Sato, Q-balls with topological charge.
  {\it Nucl Phys B} {\bf334} (1990) 489--505.

\bibitem{TW77}
J~Tjon and J~Wright, Solitons in the continuous Heisenberg
  spin chain. {\it Phys~Rev~B} {\bf15} (1977), 3470--3476.

\bibitem{KIK90}
A~M~Kosevich, B~A~Ivanov and A~S~Kovalev, Magnetic solitons.
  {\it Physics~Reports} {\bf194} (1990), 117--238.

\bibitem{PT91}
N~Papanicolaou and T~N~Tomaras, Dynamics of magnetic vortices.
    {\it Nucl~Phys~B} {\bf360} (1991), 425--462.

\bibitem{P93}
N~Papanicolaou, Dynamics of magnetic vortex rings.
 In: Singularities in Fluids, Plasmas and Optics, eds
  R~E~Caflisch \&\ G~C~Papadopoulos (Kluwer, Dordrecht, 1993).

\bibitem{C99}
N~R~Cooper, ``Smoke rings'' in ferromagnets.
  {\it Phys~Rev~Lett} {\bf82} (1999), 1554--1557.

\bibitem{W02}
R~S~Ward, Stabilizing textures with magnetic fields.
  {\it Phys~Rev~D} {\bf66} (2002) 041701.

\bibitem{W02b}
R~S~Ward, Stability of sigma-model strings and textures.
  {\it Class~Quantum~Grav} {\bf19} (2002) L17--L22.

\end{thebibliography}

\end{document}